\documentclass[prl,twocolumn,superscriptaddress]{revtex4-2}

\usepackage{graphicx}
\usepackage{dcolumn}
\usepackage{bm}
\usepackage{siunitx}
\usepackage{comment}
\usepackage[normalem]{ulem}
\usepackage[caption=false]{subfig}
\usepackage{natbib}

\begin{document}

\title{Tuning perpendicular magnetic anisotropy in ultra-low damping Li$_{0.5}$Al$_{x}$Fe$_{(2.5-x)}$O$_4$ thin films for efficient spin-orbit torque switching}

\author{Daisy O'Mahoney}
\email{daisyo@stanford.edu}
\affiliation{Department of Materials Science and Engineering, Stanford University, Stanford, California 94305, USA}
\affiliation{Geballe Laboratory for Advanced Materials, Stanford University, Stanford, California 94305, USA}

\author{Sauviz P. Alaei}
\affiliation{Department of Physics, Stanford University, Stanford, California 94305, USA}
\affiliation{Geballe Laboratory for Advanced Materials, Stanford University, Stanford, California 94305, USA}

\author{Anna Janni}
\affiliation{Department of Applied Physics, Stanford University, Stanford, California 94305, USA}
\affiliation{Geballe Laboratory for Advanced Materials, Stanford University, Stanford, California 94305, USA}

\author{Muzhda Mehrzad}
\affiliation{Geballe Laboratory for Advanced Materials, Stanford University, Stanford, California 94305, USA}

\author{Christoph Klewe}
\affiliation{Advanced Light Source, Lawrence Berkeley National Laboratory, Berkeley, California 94720, USA}

\author{Alpha T. N’Diaye}
\affiliation{Advanced Light Source, Lawrence Berkeley National Laboratory, Berkeley, California 94720, USA}

\author{Zbigniew Galazka}
\affiliation{Leibniz-Institut f\"{u}r Kristallz\"{u}chtung, Max-Born-Str. 2, 12489 Berlin, Germany}

\author{Yuri Suzuki}
\affiliation{Department of Applied Physics, Stanford University, Stanford, California 94305, USA}
\affiliation{Geballe Laboratory for Advanced Materials, Stanford University, Stanford, California 94305, USA}

\date{\today}

\begin{abstract}
Ultra-thin magnetic insulator films that simultaneously exhibit ultra-low magnon damping, perpendicular magnetic anisotropy (PMA), and low spin-orbit torque (SOT) switching current densities are highly desirable, albeit challenging, for next-generation spintronic technologies that exploit spin waves to transport information without dissipative charge currents. Here, we demonstrate this combination of properties in ferrimagnetic spinel Li$_{0.5}$Al$_{x}$Fe$_{(2.5-x)}$O$_4$ (LAFO) thin films. Through this model system, we find that PMA can be tuned by epitaxial strain in the form of chemical composition and substrate choice and that low SOT switching current densities correlate with small but finite PMA. Ultra-low damping is stabilized primarily by having only Fe$^{3+}$ as magnetically active cations with secondary effects due to increased disorder from Al substitution distribution. SOT efficiency is governed by interface quality and independent of chemical composition. By varying the Al concentration, we systematically tune the saturation magnetization and magnetic anisotropy while maintaining ultra-low Gilbert damping parameters as low as $2\times10^{-4}$ and composition-independent damping-like SOT efficiencies. We identify an optimal composition LAFO x=0.7 (Li$_{0.5}$Al$_{0.7}$Fe$_{1.8}$O$_4$), which combines ultra-low damping, stable PMA with small anisotropy fields, and low critical current densities for SOT switching, establishing it as a promising material platform for energy-efficient spin-wave and spintronic devices.
 
\end{abstract}
\maketitle

\section{Introduction}
Magnetic insulator thin films with low magnetic loss support collective spin excitations, or magnons, that can propagate coherently over long distances. The effective generation, control and transduction of these coherent excitations are essential for next generation spin-wave based devices, THz emitters and proximity-driven magnetic devices. Beyond ultra low magnetic damping, described by the Gilbert damping parameter $\alpha$, these films must exhibit: (i) isotropic in-plane magnon propagation, (ii) long spin diffusion lengths and (iii) efficient spin-to-charge interconversion. Before these materials can be incorporated into technological applications, it is necessary to understand the factors that govern magnon behavior. Depending on how magnons are generated, through thermal gradients, microwave excitations, spin-orbit torques, or optical excitations, a range of magnon modes may be excited. The scattering or coupling of these magnons with other quasi-particles may also depend on momentum. Moreover, different defects and impurities, such as vacancies and cation disorder, may systematically affect magnon behavior. 

There are a handful of magnetic insulators that can be stabilized as thin films while simultaneously meeting the requirements of PMA, isotropic in-plane magnon propagation, long spin diffusion lengths, and efficient spin-to-charge interconversion. These materials are predominantly ferrites in which the magnetism originates from the Fe$^{3+}$ ion with a 3d$^5$ configuration. The magnetic insulators identified with these properties are garnet ferrites,  spinel ferrites and hexaferrites \cite{rosenberg2018,leewong2021,Li2021,Soumah2018,Harris2012-jt}. Among the garnet ferrites engineered to exhibit PMA in thin-film form, Y$_3$Fe$_5$O$_{12}$ (YIG) has been extensively studied for spintronic applications. YIG films with PMA have demonstrated ultra-low damping values reaching as low as $\alpha\approx10^{-4}$ and have exhibited low critical SOT switching current densities on the order of 10$^6-10^7$ A/cm$^2$  \cite{Yang2024-kj}. However, these extremely low damping values can only be achieved for films grown on isotructural garnet substrates and typically require high growth temperatures (600-800$^{\circ}$C). Ultra-thin YIG films also often exhibit magnetic dead-layers at the films/substrate interface, although significant progress have been made in reducing these effects in recent years\cite{rosenberg2018, Satapathy2023-tq}. In addition to YIG, other PMA garnets with low damping and low critical switching currents have been studied, including Tm$_{3}$Fe$_{5}$O$_{12}$ (TIG), which exhibits similarly low damping values and SOT critical currents \cite{Ngaloy2025-eo,Shao2018-rn} It has been shown that these low SOT switching current densities arise from a combination of low damping and small anisotropy fields \cite{Lin2023-pp}. 

Spinel ferrites offer an alternative material platform. While they have historically been studied for bulk microwave applications, they have more recently emerged as a class of thin-film materials capable of supporting low-damping spin waves\cite{Gray2018,Riddiford2019,emori2018,zheng2020,Zheng_2023,OMahoney2023-pv,Alaei2025,Takana2025,Li2026-mh}. In particular, Li$_{0.5}$Al$_{x}$Fe$_{(2.5-x)}$O$_4$ (LAFO), derived from the parent compound Li$_{0.5}$Fe$_{2.5}$O$_4$ (LFO), has been identified as a spinel ferrite with very low damping on the order of $\alpha$ $\sim$ 10$^{-4} $\cite{Alaei2025,Zheng_2023,Takana2025,OMahoney2023-pv,Li2026-mh}. 
Aluminum substitution has been introduced into LFO (with a$_{bulk}$= 8.33 \AA) to reduce the lattice parameter and enable coherent epitaxial growth on the only commercially available spinel structure single crystal substrate MgAl$_2$O$_4$ (MAO) (with a$_{bulk}$= 8.08 \AA )\cite{Zheng_2023}. A recent study of Li$_{0.5}$Al$_{x}$Fe$_{(2.5-x)}$O$_4$ for x=0.5 and 1.0 found that increased Al substitution reduces magnetization and induces easy in-plane magnetic anisotropy due to compressive epitaxial strain imposed by the MAO substrate\cite{OMahoney2023-pv}. To stabilize PMA, LAFO x=1.0 has been grown under tensile epitaxial strain on single-crystal MgGa$_2$O$_4$ (MGO) substrates (with a$_{bulk}$=8.30 \AA) where it exhibits both PMA and low damping\cite{zheng2020}. Studies of LAFO  x=1.0 heterostructures have demonstrated efficient spin transport across Pt and Ta heavy metal bilayers, as well as low critical current densities for spin-orbit torque (SOT) switching that are more than an order of magnitude lower than those observed in comparable garnet heterostructures\cite{zheng2020,Alaei2025}. Magnon dispersion and spin diffusion lengths have also been investigated for LAFO x=1.0 films \cite{Pal2026-zi,Takana2025,Tong2026-bd,Mikhailova2026-nt}. While tensile epitaxial strain has been identified as critical to the stabilization of PMA in LAFO films, there is not a complete understanding of the interplay of damping loss and critical current density for spin-orbit torque switching in magnetic insulating thin films with PMA.

In this work, we study epitaxial LAFO thin films with Al substitution levels ranging from x=0.5 to x=1.0 on single crystal MgGa$_2$O$_4$ substrates. We demonstrate (i) how epitaxial strain, controlled through chemical composition, tunes the magnetic anisotropy from in-plane to out-of-plane, (ii) how ultra-low damping is preserved through a predominantly Fe$^{3+}$ magnetic sublattice with only secondary contributions from disorder introduced by Al substitution, (iii) how low critical SOT switching currents are achieved though weak but stable PMA with small anisotropy fields, and (iv) how SOT efficiency is governed primarily by interface quality and remains largely independent of composition. In LAFO, we identify the minimum Al concentration required to stabilize sufficient tensile epitaxial strain to induce PMA while simultaneously maintaining the lowest damping. Specifically, we find that the LAFO x=0.7 composition consistently exhibits the lowest damping reported of any PMA spinel ferrite reaching values as low as (3$\times$10$^{-4}$). The strain state of this composition on MGO induces weak PMA with very small saturation fields both in-plane and out-of-plane. In Pt/LAFO heterostructures, the epitaxial interface enables SOT switching current density values as low as 8 $\times$ 10$^6$A/cm$^2$, with switching current increasing with stronger PMA in higher Al concentration LAFO films. We further find that the damping-like SOT efficiency extracted from spin-torque ferromagnetic resonance (ST-FMR) measurements in the Pt/LAFO heterostructues is similar across all compositions for which ST-FMR could be measured. These findings demonstrate that a combination of ultra-low damping, weak magnetic anisotropy, and efficient spin transport interfaces can be achieved simultaneously, making this material system a promising candidate for future energy-efficient spintronic devices.

\begin{figure}[t]
  \centering
  {\includegraphics[width=\columnwidth]{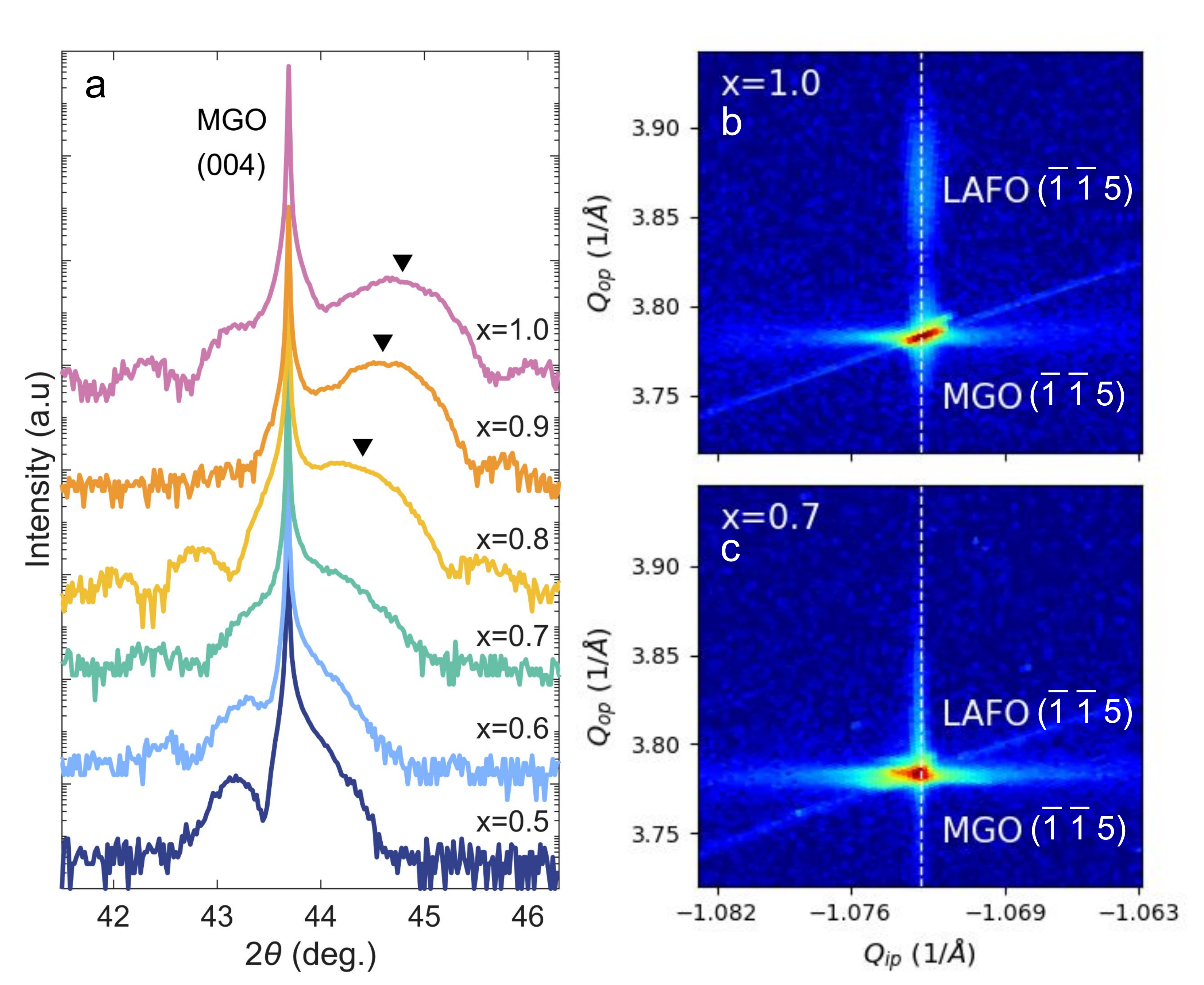}}
  \caption{Structural characterization of LAFO. (a) XRD spectra 13 nm films of the six compositions of Li$_{0.5}$Al$_{x}$Fe$_{2.5-x}$O$_4$ with x=0.5 to x=1.0. (b) Reciprocal space map of asymmetric scans about the ($\bar1\bar15$) substrate peak for of Li$_{0.5}$Al$_{0.7}$Fe$_{1.8}$O$_4$ (x=0.7). (c) Reciprocal space map of asymmetric scans about the ($\bar1\bar15$) substrate peak for of Li$_{0.5}$Al$_{1.0}$Fe$_{1.5}$O$_4$ (x=1.0)}
  \label{fig:XRD_RSM}
\end{figure}

\section{Experiment}
LAFO films were synthesized using pulsed laser deposition (PLD) on as-received single crystal (001) MgGa$_2$O$_4$ (MGO) substrates with a$_{bulk}$=8.30 \AA. MGO substrates were prepared by the Czochralski method at the Leibniz-Institut für Kristallzüchtung in Berlin, Germany. Details of this process are described elsewhere \cite{Galazka2015,Galazka2021}. Films were deposited using a 248 nm KrF laser operating at a repetition rate of 2 Hz and a fluence of 2.8 J/cm$^2$ measured at the target. Six polycrystalline targets of Li$_{0.6}$Al$_{x}$Fe$_{(2.5-x)}$O$_4$ with Al substitution ranging from x=0.5 to x=1.0 were used to create six LAFO compositions of varying Al and Fe concentrations. The targets were made by Toshima Corporation. The excess Li in the target compared to LAFO is to compensate for Li depletion during deposition. The depositions were performed in 2.0 Pa (15 mTorr) O$_2$ at substrate temperatures ranging from 400 \textdegree C (LAFO x=0.5) to 450 \textdegree C (LAFO x=1.0). LAFO films were cooled to room temperature in the chamber in an atmosphere of 13.3 Pa (100 Torr) O$_2$. Films on the order of 13 nm were used for structural and magnetic characterization while 8nm thick films were used for spin-orbit torque (SOT) measurements. 

Structural characterization of both the LAFO and Pt layers were carried out via high-resolution X-ray diffraction (XRD) and reciprocal space mapping (RSM) on a PANalytical X'Pert reflectometer. Static magnetic characterization, in the form of the field dependence of magnetization, on LAFO films was performed using a Quantum Design Evercool SQUID magnetometer. Dynamic magnetic properties of the LAFO films were characterized at room temperature using a broadband ferromagnetic resonance (FMR) system in a coplanar-waveguide geometry. In order to obtain element specific chemical and magnetic information, X-ray absorption spectroscopy (XAS) and X-ray magnetic circular dichroism measurements (XMCD) were performed at beamlines 4.0.2 and 6.3.1 of the Advanced Light Source at Lawrence Berkeley National Laboratory. Measurements were taken in magnetic fields well in excess of the saturation field of each sample at both normal and grazing incidence in luminescence yield mode.

 In order to perform spin-orbit torque (SOT) switching measurements, we fabricated Hall bars out of bilayers of LAFO and Pt. 2.5 nm of Pt was deposited on 8 nm thick LAFO films of varying Al composition using a Kurt J. Lesker (KJL) sputter system at room temperature. The Hall bars were defined by photolithography and etched by ion milling with a channel width of 10 $\mu$m and length of 40 $\mu$m.  For ST-FMR characterization of the damping-like SOT efficiency, 4 nm of Pt was deposited on LAFO also using room temperature DC sputtering. Samples were then etched into 60 $\mu$m long wires with a width of 10-20 $\mu$m using an ion mill. A coplanar waveguide of 10 nm Ti/100 nm Au was deposited on the ends of the wires using a KJL evaporator. We chose different Pt/LAFO thicknesses for SOT and ST-FMR measurements as thinner stacks improve SOT switching and thicker Pt/LAFO bi-layers improve ST-FMR signal. 

SOT switching measurements were carried out in a Quantum Design Dynacool system in fields up to 100 mT at room temperature. Spin-torque ferromagnetic resonance (STFMR) measurements were carried out at room temperature in a 2 T GMW electromagnet. RF current from an RF power source at 23 dBm was sent through the ST-FMR device and the voltage through the Pt was read using a Stanford Research SR830 lock-in amplifier. DC currents were applied using a Keithley 6221 current source attached in parallel.

\section{Results}
\subsection{Structural characterization} Aluminum substitution systematically decreases the lattice parameter of LAFO, as determined by XRD measurements of 13 nm thick films grown on MGO. Figure \ref{fig:XRD_RSM} shows a monotonic decrease in the out-of-plane lattice parameter with increasing Al concentration, indicated by the progressive shift of the LAFO (004) Bragg reflection to higher angles. Due to the close lattice match of the LAFO peaks with the substrate peak, there is a larger margin of error in the location of the lower Al composition LAFO peaks. However, for compositions of x=0.8 to x=1.0, we can easily fit the spectra using a dynamical XRD model with the in-plane lattice constant of LAFO clamped to that of the MGO substrate, as indicated by reciprocal space maps. From these fits we extract the out-of-plane lattice parameter which decreases with increasing Al composition as indicated by the black arrows in Fig \ref{fig:XRD_RSM}(a). This decrease in out-of-plane lattice parameter is consistent with increasing tensile strain in-plane. Due to the LAFO (004) peaks appearing under the MGO (004) peak for compositions x=0.5 to x=0.7, $\omega$ rocking curves were only taken for compositions x=0.8 to x=1.0. These compositions showed a typical omega rocking curve full width at half maximum (FWHM) of 0.005 $\sim$ 0.009$^{\circ}$, similar to the substrate rocking curve FWHM of 0.005 $\sim$ 0.009$^{\circ}$, indicating excellent crystallinity. (See Supplementary Section I \cite{supplemental})

Reciprocal space maps of asymmetrical peak ($\bar{1}\bar{1}5$) confirm that LAFO remains coherently strained to the MGO substrate across the full composition range. For LAFO compositions with smaller Al substitution, the film peak lies very close to the substrate peak, making it difficult to resolve the two peaks (Fig. \ref{fig:XRD_RSM}a). However, there is clearly no lateral shift of the in-plane Q$_{IP}$ of the film peak from that of the substrate peak for any of the compositions, indicating that all samples are coherently strained to MGO (Fig. \ref{fig:XRD_RSM}b and c). This is supported by previous studies on LAFO that show coherent strain on MGO  \cite{Zheng_2023, Alaei2025}. Using LAFO bulk lattice parameters which range from a$_{bulk}$ = 8.29 $\AA$ to a$_{bulk}$ = 8.20 $\AA$, we calculate biaxial tensile strain increases from approximately 0.1\% for LAFO x=0.5 to approximately 1.2\% for LAFO x=1.0, demonstrating that Al substitution provides a direct route for tuning epitaxial strain \cite{Schulkes1963,Strickler1961}. 
\begin{figure}[t]
  {\includegraphics[width=\columnwidth]{ 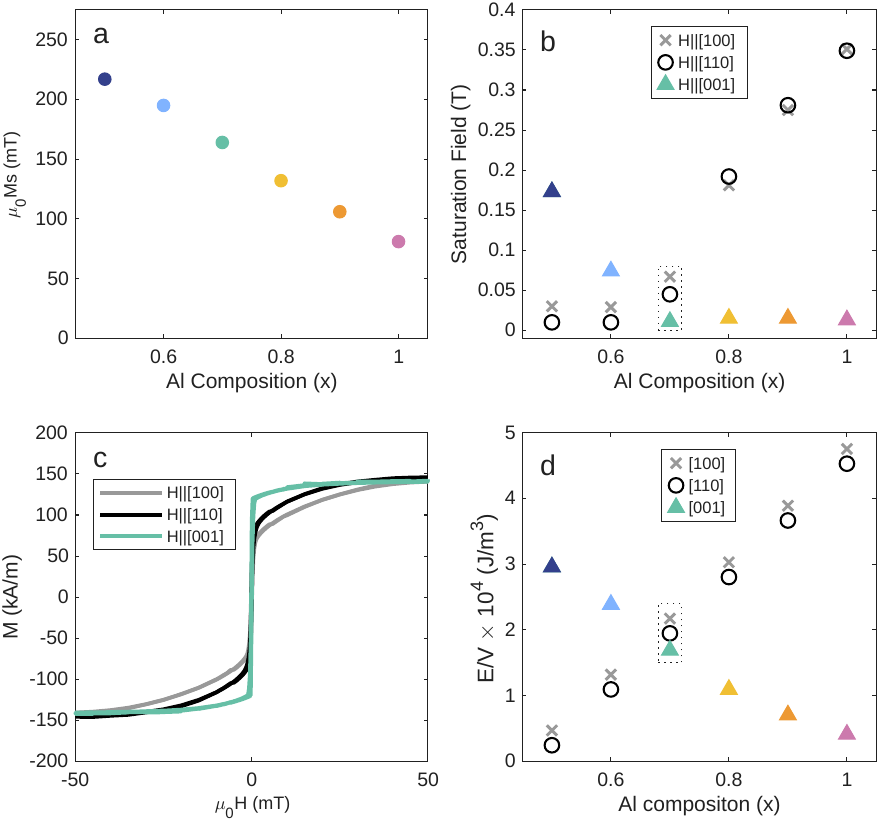}}
  \caption{Static magnetic measurements at room temperature (300 K) via SQUID magnetometry of Li$_{0.5}$Al$_{x}$Fe$_{2.5-x}$O$_4$ with x=0.5 to x=1.0. (a) Saturation magnetization (M) and (b) Saturation field ($H_s$) along the [100], [110] and [001] axes of each LAFO composition. (c) Field dependence of magnetization for LAFO x=0.7 along the in-plane [100] (gray), [110] (black) directions and out-of-plane [001] (green). (d) Calculated sum of anisotropy energies calculated for each LAFO composition using measured $M_s$ and in-plane tensile strain.} 
  \label{fig:SQUID}
\end{figure}

\subsection{Static magnetic characterization} 
Since Al substitutes for Fe in LAFO, Al composition simultaneously modifies both the saturation magnetization and the magnetic anisotropy through changes in the epitaxial strain state. SQUID magnetometry measurements performed at room temperature show that saturation magnetization decreases approximately linearly with increasing Al concentration from 200 kA/m in LAFO x=0.5 to 80 kA/m in LAFO 1.0 (Fig. \ref{fig:SQUID}(a)). Magnetization dependence of field was measured along [100], [110] and [001] directions. M$_S$ is the same along all directions for each sample within an error of $\pm$ 3-10 kA/m (See Supplementary Section IV \cite{supplemental}).

More importantly, the strain variation induced by Al substitution strongly modifies the magnetic anisotropy in LAFO, as shown in Fig. \ref{fig:SQUID}(b). The field required to saturate along the out-of-plane [001] and two in-plane directions [110] and [100] are plotted for each composition. Samples with x=0.5 and x=0.6 exhibit easy in-plane magnetic anisotropy, while samples with x$\geq$0.7 display PMA. Notably, LAFO x=0.7 exhibits weak but stable PMA with exceptionally small saturation fields in both the in-plane and out-of-plane directions of less than 70 mT, as well as coercive fields below 1 mT in all directions (Fig. \ref{fig:SQUID}(c)).

The evolution of magnetic anisotropy can be understood through a combination of magnetocrystalline, magnetoelastic and shape contributions to the anisotropy. In the absence of strain effects, bulk LAFO anisotropy constants yield easy [111], intermediate [110] and hard [100] axes. However, in thin films of LAFO on MGO, there are significant contributions from magnetoelastic and shape anisotropies. Let us assume bulk values of the magnetocrystalline anisotropy and magnetostriction constants of LFO: K$_1$=$-9.0\times10^3 J/m^3$, \ K$_2$=$2.1\times10^3 J/m^3$,  $\lambda_{100} =-2.6\times10^{-5}$ and $\lambda_{111}=2.3\times10^{-6}$ as well as a typical Young's Modulus of $Y=10^{11} J/m^3$ \cite{dionne_1969,Mazen2011}. A standard magnetostriction calculation using the measured strain and $M_s$ for each composition results in the following: although all samples are under tensile strain, the magnetoelastic contribution is not enough to overcome shape and magnetocrystalline anisotropy below x=0.7 with the easy axis along [110] and intermediate and hard directions along [100] and [001] respectively (Fig. \ref{fig:SQUID}(d)). At LAFO x=0.7, the easy direction switches to [001] with a magnetically hard in-plane direction and the stabilization of PMA.
These results demonstrate that epitaxial strain tuning through Al substitution provides a direct route for controlling the magnetic anisotropy in these films. The calculated anisotropy behavior agrees closely with the measured SQUID data (Fig. \ref{fig:SQUID}(b)). Details of these calculations can be found in the Supplementary Section II \cite{supplemental}. 

\begin{figure}[t]
  {\includegraphics[width=\columnwidth]{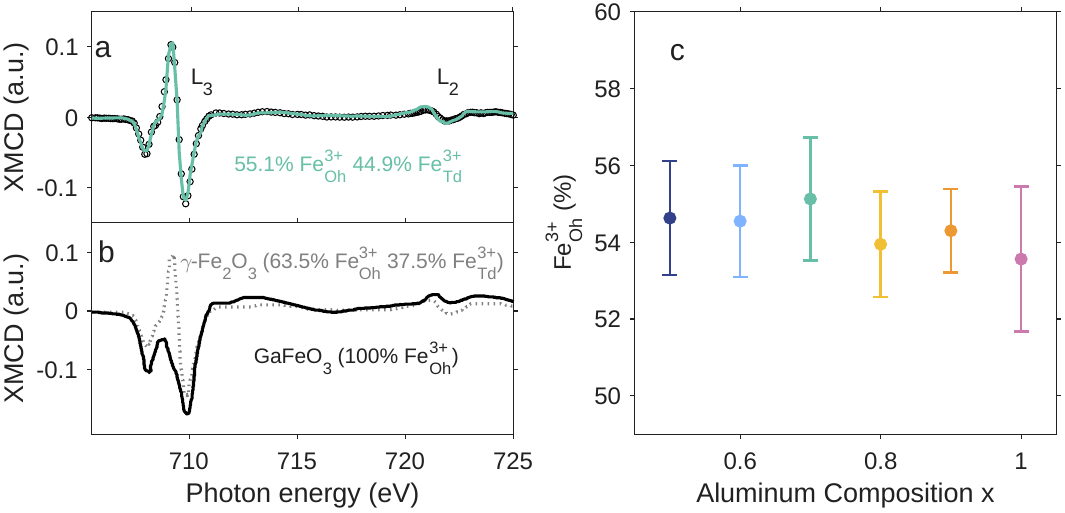}}
  \caption{X-ray magnetic circular dichroism (XMCD) of Fe L$_2,3$ edges in LAFO (a) XMCD of LAFO x=0.7 (black) and fit to reference spectra (green) (b) Reference XMCD spectra used to fit (a) of $\gamma$-Fe$_2$O$_3$ (62.5\% Fe$^{3+}_{Oh}$ 37.5\% Fe$^{3+}_{Td}$) \cite{brice-profeta_2005} and GaFeO$_3$ (100\% Fe$^{3+}_{Oh}$) \cite{Kim2006} (c) Percent of Fe$^{3+}$ occupying the octahedral sites in each LAFO composition.} 
  \label{fig:XMCD}
\end{figure} 

XMCD measurements indicate that the magnetically active ions are Fe$^{3+}$ ions. XAS spectra were taken at the L$_2$ and L$_3$ peaks of Fe for each of the LAFO compositions. XMCD spectra is created by taking the difference in the x-ray absorption spectra (XAS) of left and right circularly polarized light so that the measured net magnetization is parallel and antiparallel to the direction of the X-ray polarization. The XMCD spectra was then fit using two reference spectra of known Fe$^{3+}$ octahedral and tetrahedral distribution: $\gamma$-Fe$_2$O$_3$ (62.5\% Fe$^{3+}_{Oh}$ 37.5\% Fe$^{3+}_{Td}$) \cite{brice-profeta_2005} and GaFeO$_3$ (100\% Fe$^{3+}_{Oh}$) \cite{Kim2006}. We are able to fit spectra for all compositions with only Fe$^{3+}$ and no Fe$^{2+}$ as has been the case for previous LAFO studies \cite{OMahoney2023-pv,emori2018,Takana2025}. Fig. \ref{fig:XMCD}(a) show the XMCD spectra of Fe from LAFO x=0.7 which is representative of the typical spectra and fits for each composition. XAS and XMCD spectra for all compositions can be found  in Supplementary Section III \cite{supplemental}. From our fits, we extract an Fe$^{3+}$ distribution of 54\% +/- 2\% Oh and 46\% +/- 2\% Td for each composition, indicating that within this composition range, the Al does not noticeably affect the Fe$^{3+}$ distribution.

\begin{figure}[t]
  {\includegraphics[width=\columnwidth]{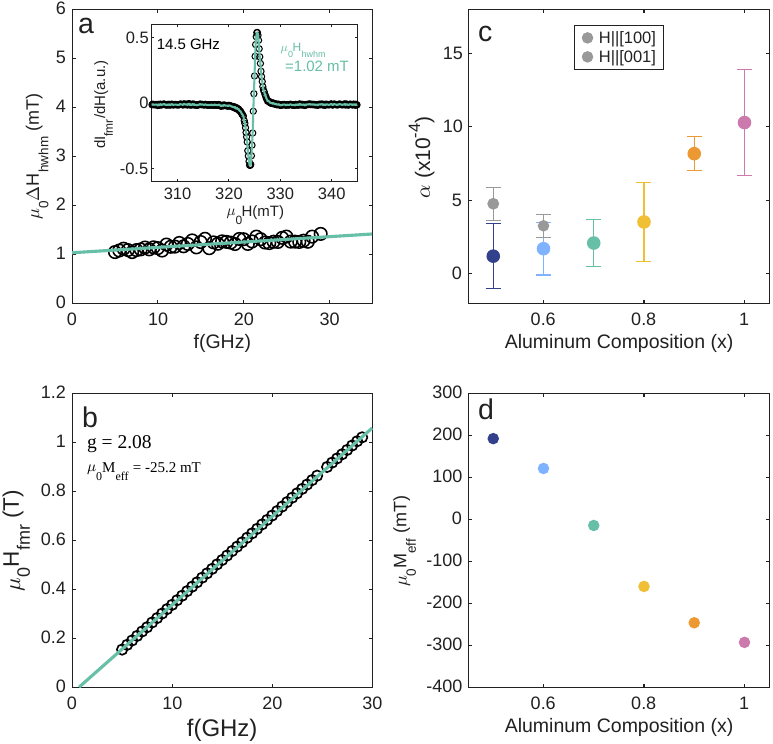}}
  \caption{Ferromagnetic resonance of LAFO. (a) frequency dependence of half-width-half-maximum linewidth ($\mu_0$H$_{\text{hwhm}}$) of a 13 nm sample of Li$_{0.5}$Al$_{0.7}$Fe$_{1.8}$O$_4$ (x=0.7), with the green line as a fit to (Eq. 1). The inset shows a sample FMR spectra at 14.5 GHz. for LAFO x=0.7. (b) FMR resonance field $\mu_0$H$_\text{fmr}$ as a function of frequency with field along [001] for LAFO x=0.7. The green line indicates a fit to the Kittel equation (Eq. 2). (c) Damping parameter of LAFO x=0.5 to x=1.0 taken with field along the easy plane [100] for x=0.5 and x=0.6 and [001] for x=0.7 to x=1.0}
  \label{fig:FMR}
\end{figure}

 \subsection{Ferromagnetic resonance} 
 Broadband FMR measurements reveal ultra-low magnetic damping across the entire composition range. A typical FMR spectra for LAFO x=0.7 is shown in the inset of Fig. \ref{fig:FMR}a with a linewidth of 1.02 mT (10.2 Oe). In order to extract the Gilbert damping parameter which is a measure of magnetic loss, the FMR spectra is taken at frequencies from 5 GHz to 30 GHz and the width of the absorption peak is plotted vs frequency and fit to: $\Delta H_{\text{hwhm}}=\Delta H_0 + \alpha\frac{h}{g \mu_B \mu_0} f$ (Eq. 1), where $\Delta H_0$ is the inhomogeneous broadening attributed to disorder and defects, $\alpha$ is the Gilbert damping parameter, $h$ is Planck's constant, $g$ is the Lande g-factor (quantified from Fig. \ref{fig:FMR}(b)), $\mu_B$ is the Bohr magneton and $\mu_0$ is the permeability of free space. Fig. \ref{fig:FMR}(a) shows the frequency dependence of the FMR linewidth for LAFO x=0.7 with an ultra-low damping parameter of $\alpha=3.06 \times 10^{-4}$. 
 
We extract the Gilbert damping parameter for each composition (Fig. \ref{fig:SQUID}(c)) and average over at least two samples for each composition. When measuring FMR with field parallel to the [001] direction across compositions, we see a minimum in the Gilbert damping parameter at LAFO x=0.7. This is because the [001] direction becomes the magnetically hard axis for the in-plane compositions LAFO x=0.5 and x=0.6 while it is the magnetically easy direction for higher Al doping. When we measure damping parameters measured with the field along easy directions across compositions (i.e., along the easy plane $H\parallel [100] $ for the in-plane compositions and $H\parallel [001] $ for PMA samples), there is a clear increase in $\alpha$ with increasing Al that is attributed to increased disorder from Al doping.  The lower Al composition samples have $\alpha$$\sim$1.5 $\times$ 10$^{-4}$ to 3.0$ \times 10^{-4}$ which is the lowest damping parameter reported in any spinel ferrites \cite{emori2018,OMahoney2023-pv,Riddiford2019,Zheng_2023,zheng2020,emori2017,Gray2018}, comparable to the lowest damping parameters seen in thin films of YIG \cite{Schmidt2020,Serha2026-te,Hauser2016,Soumah2018}.

Frequency dependence of the FMR field allows us to deduce the spectroscopic g factor as well as  the effective magnetization $M_{eff}$ from the Kittel equation: $f= \frac{g \mu_B\mu_0}{h} (H_\text{fmr} - M_\text{eff}) \ $  (Eq. 2), where $M_\text{eff}=M_\text{s}-H_{2,\perp}$, and H$_{2,\perp}$ is the out-of-plane uniaxial anisotropy field\cite{emori2018, farle_1998, lindner2009}. H$_{2,\perp}$ is positive for all samples and increases with increasing Al composition. At an Al composition x=0.7, H$_{2,\perp}$ is large enough to overcome $M_\text{s}$, causing $M_\text{eff}$ to become negative (Fig. \ref{fig:FMR}d) and hence exhibit PMA. The negative $M_\text{eff}$ and PMA in all compositions above x=0.7 is in agreement with static magnetic measurements (Fig. \ref{fig:SQUID}). The average g for all LAFO compositions was found to be 2.04 +/- 0.02 and consistent with a total orbital angular momentum of L=0 for Fe$^{3+}$. This value indicates relatively weak spin-orbit coupling, which is favorable for suppressing magnetic damping and maintaining coherent magnon transport. The extracted g-factor is also consistent with the XMCD measurements, which show that the magnetism originates overwhelmingly from Fe$^{3+}$ ions with L=0. 

\subsection{Spin torque ferromagnetic resonance}
To electrically modulate LAFO magnetization and quantify the efficiency of this process, we performed spin-torque ferromagnetic resonance (ST-FMR) measurements. By fabricating Pt/LAFO bilayers, we are able to extract the spin-orbit torque (SOT) efficiency, which  quantifies spin-to-charge interconversion and incorporates contributions from both the spin Hall angle of the Pt and the spin transparency of the Pt/LAFO interface. In ST-FMR measurements, a microwave current is passed in-plane through the Pt layer, which generates an anti-damping torque on the LAFO magnetization via the spin Hall effect. Under appropriate magnetic field conditions, these torques drive coherent magnon precession in the LAFO, which in turn produces oscillations in the Pt resistance through spin Hall and anisotropic magnetoresistance effects\cite{Shao2021,Schreier2015-cq,Sklenar2015-fg}. The interference between the oscillating resistance and microwave current generates a dc voltage signal. The measured voltage spectra can be fit using a superposition of symmetric and antisymmetric Lorentzian functions to extract the half-width-at-half-maximum linewidth $\Delta$H. An additional dc current can be applied alongside the microwave current to determine the SOT efficiency\cite{Karimeddiny2021-mo,Li2021-lf}. The applied dc current modifies the FMR linewidth linearly, allowing the damping-like torque efficiency to be extracted \cite{Liu2011-vg,Nan2015-dq}. Fitting details are provided in Supplementary Section V \cite{supplemental}.

\begin{figure}[t]
	\centering
  {\includegraphics[width=\columnwidth]{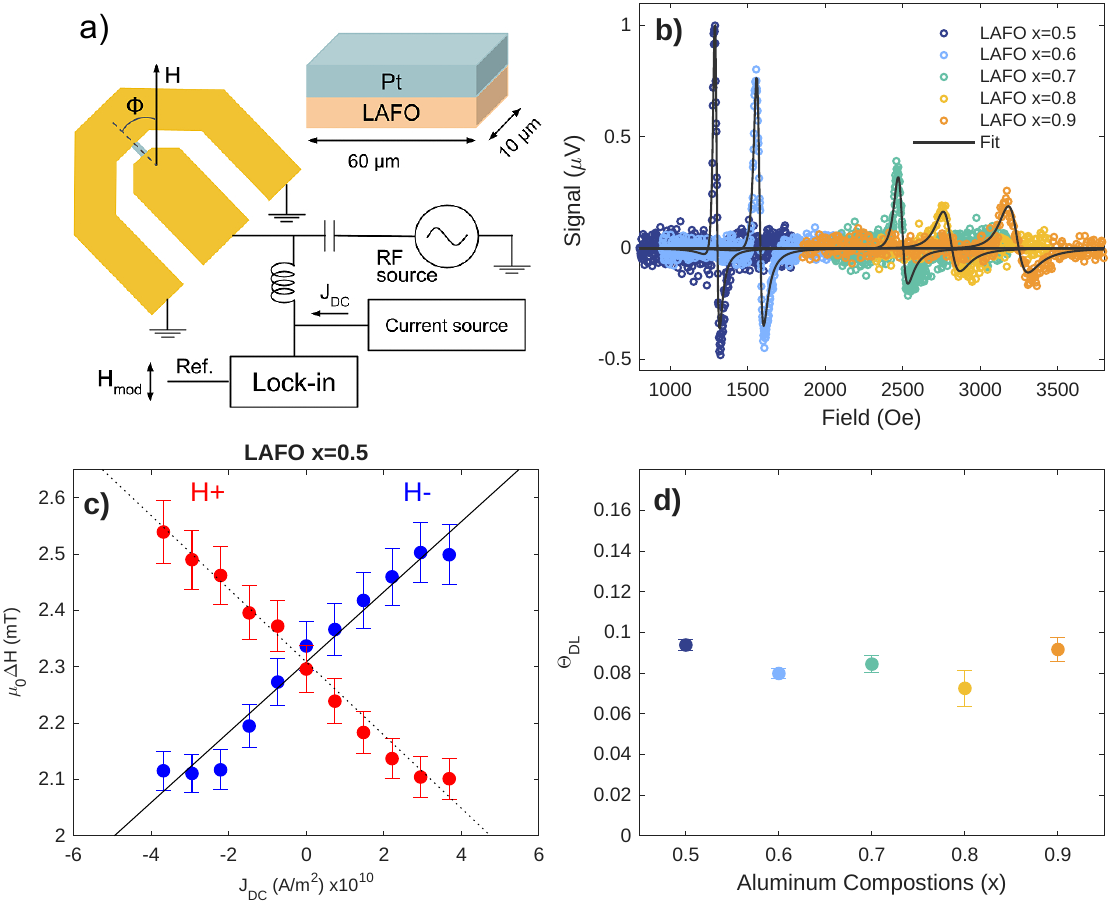}}
  \caption{Spin-torque ferromagnetic resonance (a) Pt/LAFO stack etched to a 60 $\mu$m × 10 $\mu$m strip covered with an Au electrode and connected to external circuit. (b) FMR resonance signals at 5.5 GHz for each composition with total Lorentzian fits in black (c) Dependence of linewidth $\Delta$H on dc bias current density J$_{dc}$ for LAFO x=0.5 (13 nm)/Pt(4 nm). Linewidths and linear fits under positive (blue circles and dashed line) and negative (red circles and solid line) magnetic fields (d) $\theta_{DL}$ for all Pt/LAFO of all LAFO compositions}
  \label{fig:STFMR}
\end{figure}

Due to the required field and current geometry, ST-FMR is fundamentally an in-plane measurement, making it challenging to implement in PMA magnetic systems. In most PMA materials, the in-plane damping is relatively large, resulting in broad, low-amplitude resonance peaks that complicate accurate extraction of SOT efficiencies. Figure \ref{fig:STFMR}(b) shows that LAFO samples with in-plane magnetic anisotropy exhibit sharper and higher-amplitude resonance peaks compared to the PMA compositions. However, the exceptionally small in-plane saturation fields of LAFO x=0.7 enables relatively strong ST-FMR signals despite the presence of PMA. This unique combination of weak but stable PMA and ultra-low damping allows spin-torques characterization in PMA LAFO compositions (x=0.7, 0.8 and 0.9) that would otherwise be difficult to access using conventional ST-FMR measurements (See Supplementary Section VI \cite{supplemental}). We extract damping-like SOT efficiencies in the range of 0.073 to 0.094 with a mean of 0.084 for LAFO x=0.5, 0.6, 0.7, 0.8 and 0.9, indicating consistently efficient spin-to-charge interconversion across the full compositional range. Importantly, the relatively constant SOT efficiency across compositions suggests that the Pt/LAFO interface transparency remains the same despite substantial changes in magnetic anisotropy and strain state \cite{Pai2015-lc}.  

\subsection {Spin-orbit torque switching}
Current-induced SOT switching of the magnetic state of LAFO films has also been demonstrated in our Pt/LAFO bilayers across two compositions. PMA magnetic films provide for faster switching and higher bit densities in applications and therefore are more promising compared to their in-plane counterparts. The tunable PMA realized in LAFO films provides a unique platform for understanding the factors governing critical switching current density in magnetic insulator heterostructures. In SOT switching, electrical current in the Pt layer generates a spin current via the spin Hall effect, which applies an anti-damping torque to the LAFO magnetization. Under appropriate conditions, this torque can switch the magnetization direction of the LAFO. A number of factors are known to promote efficient SOT switching, including low magnetic damping, small anisotropy barriers, and efficient spin-transparent interfaces \cite{Shao2021,Lin2023-pp}. Since our heterostructures with varying LAFO composition exhibit  consistent interfacial quality and nearly composition-independent SOT efficiencies, we are able to isolate the influence of PMA strength and magnetic anisotropy on critical switching current density. 

\begin{figure}[t]
	\centering
  {\includegraphics[width=\columnwidth]{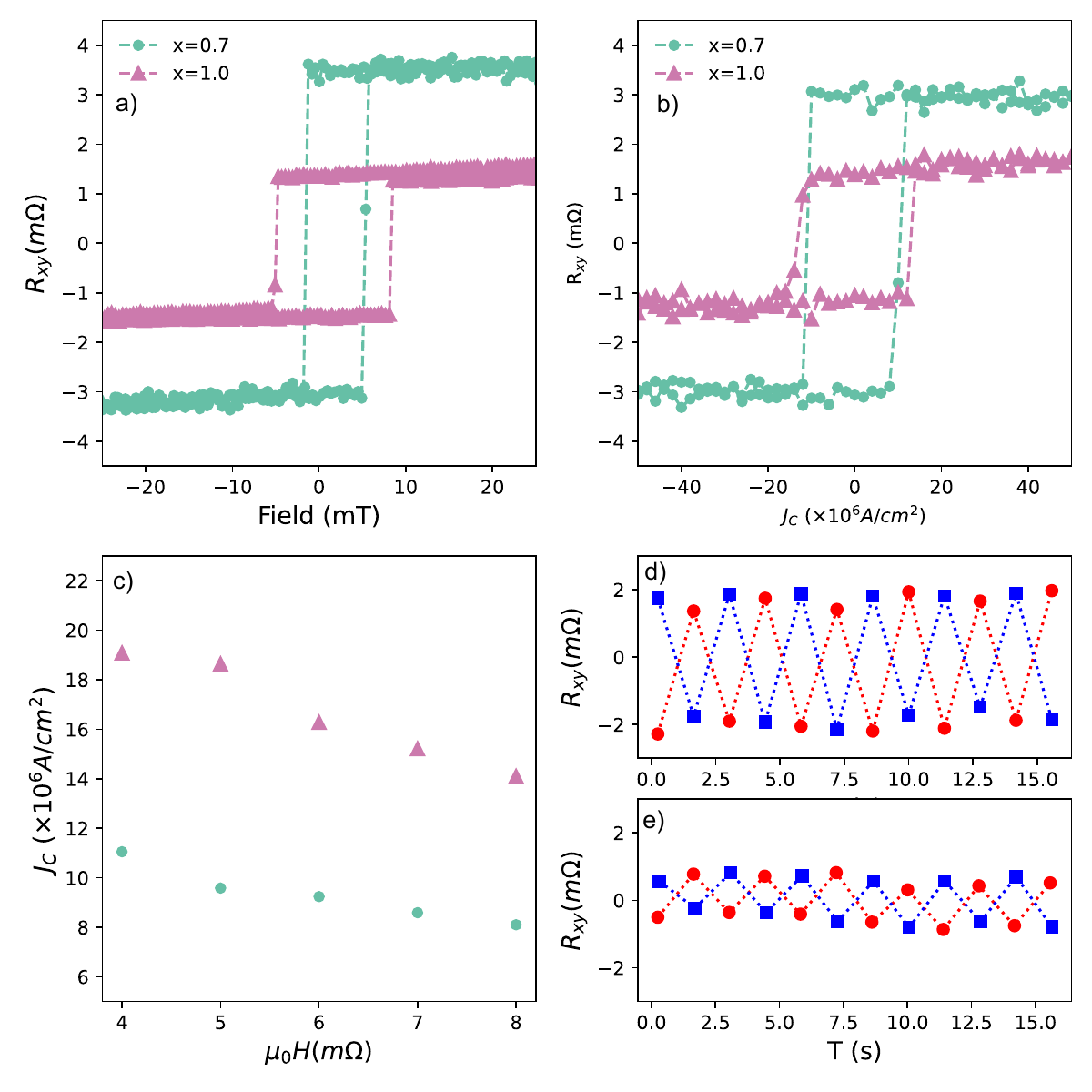}}
  \caption{Current induced spin-orbit torque switching of the magnetization in two LAFO compositions with PMA: x=0.7 (green) and x=1.0 (pink). (a) Hall(R$_{xy}$) resistance as a function of out-of-plane field showing clear hysteresis due to the anomalous Hall effect in the Pt from the magnetization of the adjacent LAFO layer for both compositions. (b) R$_{xy}$ measured as a function of current density with applied field Hx of 7 mT along the current direction. (c) Critical switching current density J$_C$ as a function of in-plane field. (d-e) R$_{xy}$ measured after current pulses of alternating sign for applied field H$_x$ of +/- 5 mT  (red/blue) along the current direction for (d) LAFO x=0.7 and LAFO x=1.0 (d) }
  \label{fig:SOT}
\end{figure}

We focus here on heterostructures containing PMA LAFO, particularly LAFO x=0.7 with weakest PMA and LAFO x=1.0 with strongest PMA. We first measure Hall resistance R$_{xy}$ as a function of applied out-of-plane field H$_z$ after subtracting the linear ordinary Hall contribution. Due to the magnetic proximity effect induced by LAFO, Hall transport measurements in the Pt layer exhibit an anomalous Hall effect that reflects the magnetic state of the LAFO film\cite{Chen2013-lj}. We observe larger magnetic switching fields in the Hall resistance hysteresis loops of LAFO x=1.0 compared to LAFO x=0.7, consistent with larger coercive fields observed in the magnetization loops of the LAFO x=1.0 bilayer (Fig. \ref{fig:SQUID}d). For SOT switching measurements, 5 ms dc current pulses are applied along the Hall bar and the Hall resistance R$_{xy}$ is measured after each pulse. Charge current flowing through Pt generates a spin current via the spin Hall effect that propagates towards the Pt/LAFO interface. The orientation of spin polarization with respect to the LAFO magnetization determines whether spins are absorbed or reflected at the interface, thereby modulating the measured resistance. As a result, plots of Hall resistance as a function of applied current exhibit clear hysteretic behavior. The difference in Hall signal amplitude between the two compositions reflects the different saturation magnetization values of the LAFO layers. We then measure R$_{xy}$ as a function of current pulse density $J_{DC}$ to probe the hysteretic behavior associated with electrically induced magnetization switching. To break the symmetry during the SOT measurements, a small in-plane magnetic helper field is applied. The current-dependent hysteresis loops closely resemble those obtained during magnetic field sweeps, confirming current-induced SOT switching of the LAFO layer(Fig. \ref{fig:SOT}b). To demonstrate repeatable switching, we measure R$_{xy}$ following a sequence of alternating positive and negative dc current pulses under alternating in-plane magnetic fields (Fig. \ref{fig:SOT}c). We observe consistent and deterministic switching for both compositions.

We further measure the critical switching current density as a function of the in-plane helper field H$_x$ in both compositions (Fig. \ref{fig:SOT}d). As expected, increasing the magnitude of the helper field reduces the critical switching current density. Notably, LAFO x=0.7 exhibits significantly lower switching current densities and larger Hall signal amplitudes compared to LAFO x=1.0. This behavior is consistent with the substantially  smaller out-of-plane saturation field and weaker anisotropy barrier of LAFO x=0.7, which reduces the current required for magnetization reversal. These results demonstrate that weak but stable PMA provides an optimal regime for minimizing switching current while preserving perpendicular magnetic order. As was previously observed in thin films of LAFO x=1.0, the critical switching current is on the order of those reported in YIG heterostructures (typically 10$^6$-10$^7$ A/cm$^2$) \cite{Fedel2026-mk, Guo2019-wo,Shiota2026-uv,Yang2024-kj}. In several systems, the helper field has been eliminated through device geometry engineering or introduction of an exchange bias \cite{Oh2016-tx,Chen2019-al,Yang2024-kj} and these approaches will be explored in further studies. The nearly composition-independent damping-like SOT efficiency further indicates that the reduction in switching current density primarily originates from tuning intrinsic magnetic properties—namely damping and anisotropy barriers—rather than changes in interfacial spin transparency.

\section{Discussion}
Our LAFO films and bilayers demonstrate that PMA can be systematically tuned while maintaining ultra-low magnetic damping, consistent SOT efficiencies, and efficient spin-transparent interfaces. Increased Al substitution does introduce additional disorder, as evidenced by the increased inhomogeneous broadening observed in the frequency dependence of the FMR linewidth. However, the Gilbert damping increases only negligibly for compositions with x $<$ 0.9. Furthermore, increased Al substitution does not significantly alter the Fe$^{3+}$ distribution among cation sites. Previous studies have shown that  Pt grows epitaxially on LAFO \cite{Zheng_2023}. This growth produces highly ordered Pt/LAFO interfaces that maintain efficient transport and nearly composition-independent SOT efficiencies across the full Al composition range. 

With ultra-low magnetic damping, consistent SOT efficiencies, and high-quality epitaxial interfaces in Pt/LAFO bilayers, we find that Al substitution provides the necessary tuning of the magnetic anisotropy and saturation magnetization to reveal the factors governing effective current-induced SOT switching. Increased Al concentration in LAFO simultaneously decreases the saturation magnetization and lattice parameter,  driving the evolution of the magnetic anisotropy from in-plane to PMA through increasing  tensile epitaxial strain on MGO substrates. Importantly, this strain-driven tuning enables access to a weak-PMA regime that minimizes anisotropy barriers while preserving perpendicular magnetic stability. By maintaining a finite but stable PMA together with sufficiently large magnetization, we demonstrate that LAFO x=0.7/Pt bilayers exhibit the lowest critical switching current densities, establishing this composition as an optimized regime for low-power magnetic-insulating spintronics. 

\section{Conclusions}
In conclusion, we demonstrate that a weak but stable PMA combined with a relatively large saturation magnetization gives rise to the most efficient SOT switching in Pt/LAFO bilayers. Al substitution of Fe in LAFO films on MGO substrates systematically modulates the saturation magnetization and magnetic anisotropy through epitaxial strain tuning while maintaining ultra-low Gilbert damping parameters and nearly composition independent SOT efficiencies. Through a detailed investigation of the static and dynamic magnetic properties, as well as spin-to-charge interconversion in these heterostructures, we identify the factors governing efficient SOT switching: ultra-low damping, small anisotropy barriers and efficient spin-transport interfaces. Together these results are promising for next generation spintronics based on spin wave transport.

\section{Data Availability}
All data supporting the findings of this article is openly available at https://doi.org/10.25740/np409xc4374.

\section*{Acknowledgments}
This work was supported by the U.S. Department of Energy, Director, Office of Science, Office of Basic Energy Sciences, Division of Materials Sciences and Engineering under Contract No. DESC0008505. S.A., who contributed to SOT switching experiments, was supported by the Air Force Office of Scientific Research, Grants No. FA9550-23-1-0344. Part of this work was performed at nano@stanford RRID:SCR\_026695. This research used resources of the Advanced Light Source, which is a DOE Office of Science User Facility under contract no. DE-AC02-05CH11231.
The authors thank Jutta Schwarzkopf from Leibniz-Institut für Kristallzüchtung for critical reading of the paper.
\bibliography{refs}

\end{document}


\title{Supplementary Information: Tuning perpendicular magnetic anisotropy in ultra-low damping Li$_{0.5}$Al$_{x}$Fe$_{(2.5-x)}$O$_4$ thin films for efficient spin-orbit torque switching} 

\author{Daisy O'Mahoney}
\affiliation{ Department of Materials Science and Engineering, Stanford University, Stanford, California 94305, USA}
\affiliation{Geballe Laboratory for Advanced Materials, Stanford University, Stanford, California 94305, USA}

\author{Sauviz P. Alaei}
\affiliation{Department of Physics, Stanford University, Stanford, California 94305, USA}
\affiliation{Geballe Laboratory for Advanced Materials, Stanford University, Stanford, California 94305, USA}

\author{Anna Janni}
\affiliation{Department of Physics, Stanford University, Stanford, California 94305, USA}
\affiliation{Geballe Laboratory for Advanced Materials, Stanford University, Stanford, California 94305, USA}

\author{Muzhda Mehrzad}
\affiliation{Department of Electrical Engineering and Computer Sciences, University of California, Berkeley, Berkeley, CA 94720, USA}

\author{Christoph Klewe}
\affiliation{Advanced Light Source, Lawrence Berkeley National Laboratory, Berkeley, California 94720, USA}

\author{Alpha T. N’Diaye}
\affiliation{Advanced Light Source, Lawrence Berkeley National Laboratory, Berkeley, California 94720, USA}

\author{Zbigniew Galazka}
\affiliation{Leibniz-Institut f\"{u}r Kristallz\"{u}chtung, Max-Born-Str. 2, 12489 Berlin, Germany}

\author{Yuri Suzuki}
\affiliation{Department of Applied Physics, Stanford University, Stanford, California 94305, USA}
\affiliation{Geballe Laboratory for Advanced Materials, Stanford University, Stanford, California 94305, USA}

\maketitle

\section{Structural Characterization}
\begin{figure}[t]
    \centering
    \renewcommand{\thefigure}{S\arabic{figure}}
    \includegraphics[width=\columnwidth]{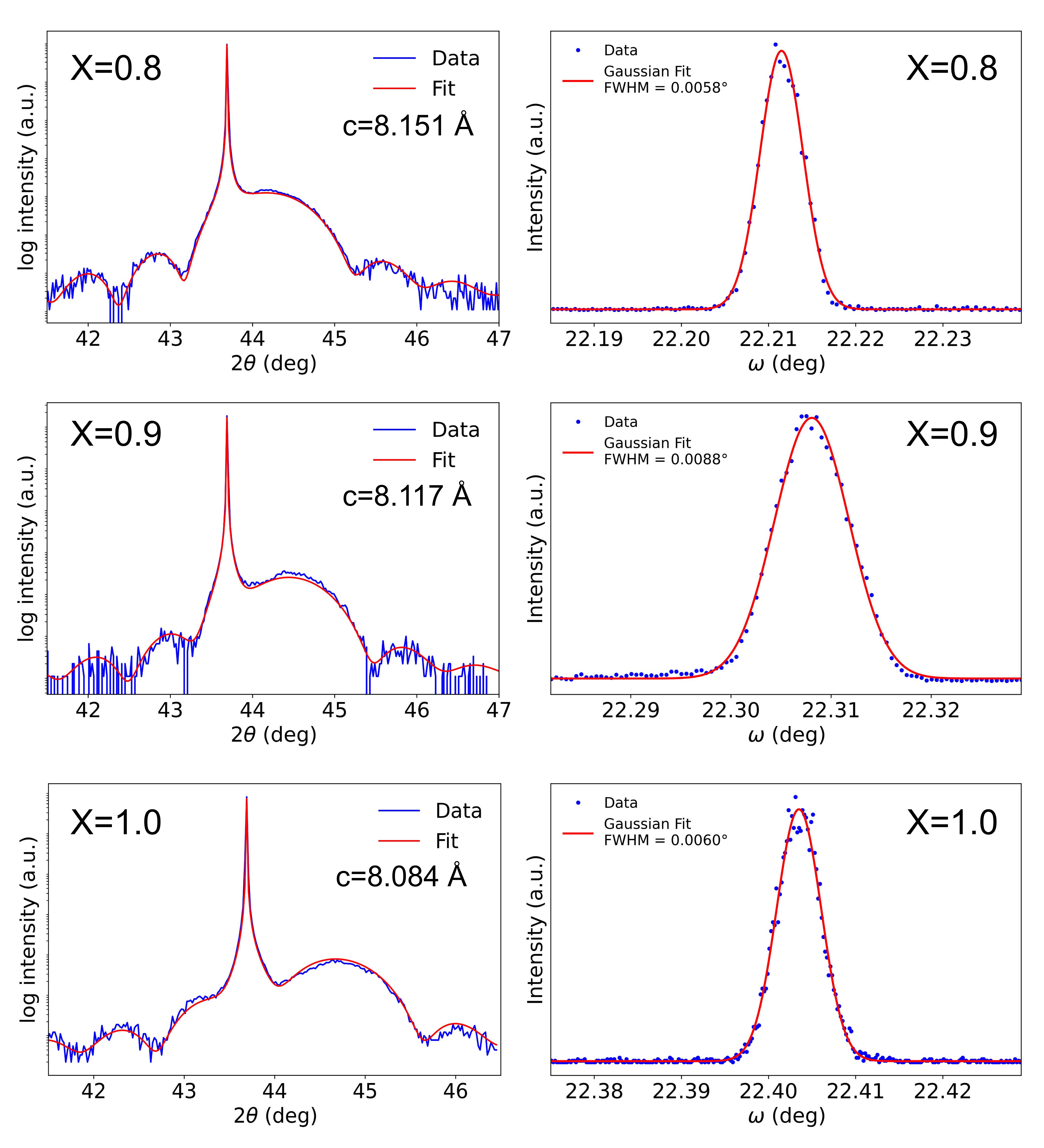}
    \caption{Fitted XRD data for compositions x=0.8 to x=1.0. The left column shows symmetrical 2$\theta-\omega$ scans of the (004) peak with data shown in blue and the fit using a dynamical x-ray model in red. The right column shows corresponding $\omega$ scans of the (004) LAFO peaks where the data is shown in blue and a Gaussian fit is shown in red. $\omega$ FWHM values range from 0.0056\textdegree to 0.0088\textdegree}
    \label{fig:XRD fitting}
\end{figure}

\begin{figure}[t]
    \centering
    \renewcommand{\thefigure}{S\arabic{figure}}
    \includegraphics[width=0.8\columnwidth]{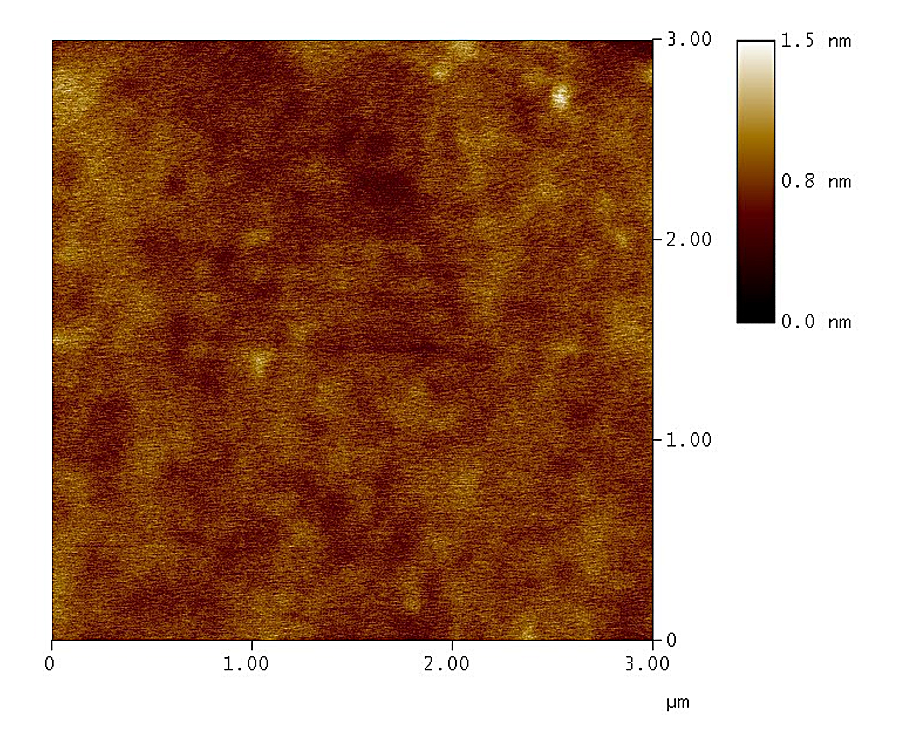}
    \caption{Atomic force microscopy of LAFO x=0.7 measured directly after deposition shows a very smooth surface with RMS roughness of 0.11 nm.}
    \label{fig:AFM}
\end{figure}

X-ray diffraction symmetric scans around the (004)
Bragg peaks of LAFO x=0.8 to x=1.0 are shown in Fig \ref{fig:XRD fitting}. These compositions have clearly resolvable peaks and are fit using a dynamical x-ray model to extract the out-of-plane lattice parameters. The lattice parameters decrease from 8.151 \AA in LAFO x=0.8 to 8.084 \AA in LAFO x=1.0 with an error of $\pm 0.002$ \AA. $\omega$ rocking curve scans of these three compositions and fit to a Gaussian function (Fig \ref{fig:XRD fitting}). The FWHM of the omega peaks range from 0.0056\textdegree\ to 0.0088\textdegree\ indicating excellent crystallinity. An atomic force microscopy image was taken of LAFO x=0.7 directly after deposition (Fig \ref{fig:AFM}). The surface is exceptionally flat with an RMS roughness of 0.11 nm.  

\section{Strain Calculations}
We performed magnetic anisotropy calculations by taking into account  magnetocrystalline anisotropy $E_{mc}$ (Eqn. S1), magnetoelastic anisotropy $E_{me}$ (Eqn. S2) and shape anisotropy $E_{s}$ contributions (Eqn. S3). In these calculations we assume that LAFO has approximately the anisotropy constants and Young's Modulus as found in bulk LFO and those were used for all compositions. For magnetocrystalline anisotropy:

\begin{equation}
\renewcommand{\theequation}{S\arabic{equation}}
E_{mc}=K_0+K_1(\alpha_1^2\alpha_2^2+\alpha_2^2\alpha_3^2+\alpha_3^2\alpha_1^2)+K_2(\alpha_1^2\alpha_2^2\alpha_3^2) +...
\end{equation}, 

we assume $K_0=0$, $K_1=-9.0\times10^3 J/m^3$ and $K_2=2.1\times10^3J/m^3$ and ignore all higher order anisotropy constants \cite{dionne_1969}. $\alpha_1$, $\alpha_2$ and $\alpha_3$ represent the cosine of the angle between the magnetic moment and the crystallographic axis directions a, b and c respectively. Magnetoelastic anisotropy energy can be written as:

\begin{equation}
\renewcommand{\theequation}{S\arabic{equation}}
E_{me}=-\frac{3}{2}\lambda_{100}\sigma(\alpha_1^2\gamma_1^2+\alpha_2^2\gamma_2^2+\alpha_3^2\gamma_3^2)-3\lambda_{111}\sigma(\alpha_1\alpha_2\gamma_1\gamma_2+\alpha_2\alpha_2\gamma_2\gamma_3+\alpha_3\alpha_1\gamma_3\gamma_1)
\end{equation}

where $\gamma_1, \gamma_2,\gamma_3$ are the directional cosines of the applied stress $\sigma$ and $\lambda_{100}$ and $\lambda_{111}$ are the magnetostrictive constants along the [100] and [111] directions respectively. We assume bulk LFO magnetostrictive constants $\lambda_{100}=-2.6\times10^{-5}$ $\lambda_{111} = 2.3 \times 10^{-6}$ \cite{dionne_1969}. To calculate the stress, we assume a Young's modulus of bulk LFO Y $\sim 10^{11}$ J/m$^3$ \cite{Mazen2011} and calculate the strain from comparing lattice constants obtained from RSM data to bulk lattice constants \cite{Shiota2026-uv,Strickler1961}. Shape anisotropy of a thin film is approximated by:
\begin{equation}
\renewcommand{\theequation}{S\arabic{equation}}
E_{s}=\frac{1}{2}\mu_0M_s^2cos^2\theta
\end{equation}
where $\theta$ is the angle between the film normal and the magnetization direction and M$_s$ is the saturation magnetization. The total anisotropy energy is then calculated for each composition along the crystallographic directions [100], [110] and [001] by summing each of the above anisotropy terms. Table S1 summarizes the anisotropy energies for each composition. 

\begin{table}[t]
\renewcommand{\thetable}{S\arabic{table}}
\caption{Strain calculation results and bulk lattice parameters}
\label{tab:params}
\centering
\small
\setlength{\tabcolsep}{12pt}
\begin{tabular}{cccccc}
\toprule
\makecell{Al \\ Comp.} &
\makecell{Bulk Lattice \\ Parameter (\AA)} &
\makecell{$\varepsilon$ \\ (\%)} & 
\makecell{$E_{[100]}$ \\ ($10^4$ J/m$^3$)} &
\makecell{$E_{[110]}$ \\ ($10^4$ J/m$^3$)} &
\makecell{$E_{[001]}$ \\ ($10^4$ J/m$^3$)} \\
\midrule
0.5 & 8.29 & 0.12 & 0.47 & 0.25 & 2.96 \\
0.6 & 8.28 & 0.34 & 1.32 & 1.10 & 2.39 \\
0.7 & 8.26 & 0.56 & 2.17 & 1.95 & 1.69 \\
0.8 & 8.24 & 0.78 & 3.03 & 2.81 & 1.09 \\
0.9 & 8.22 & 1.00 & 3.89 & 3.67 & 0.71 \\
1.0 & 8.20 & 1.22 & 4.76 & 4.53 & 0.41 \\
\bottomrule
\end{tabular}
\end{table}

\begin{figure}[t]
    \renewcommand{\thefigure}{S\arabic{figure}}
    \includegraphics[width=\columnwidth]{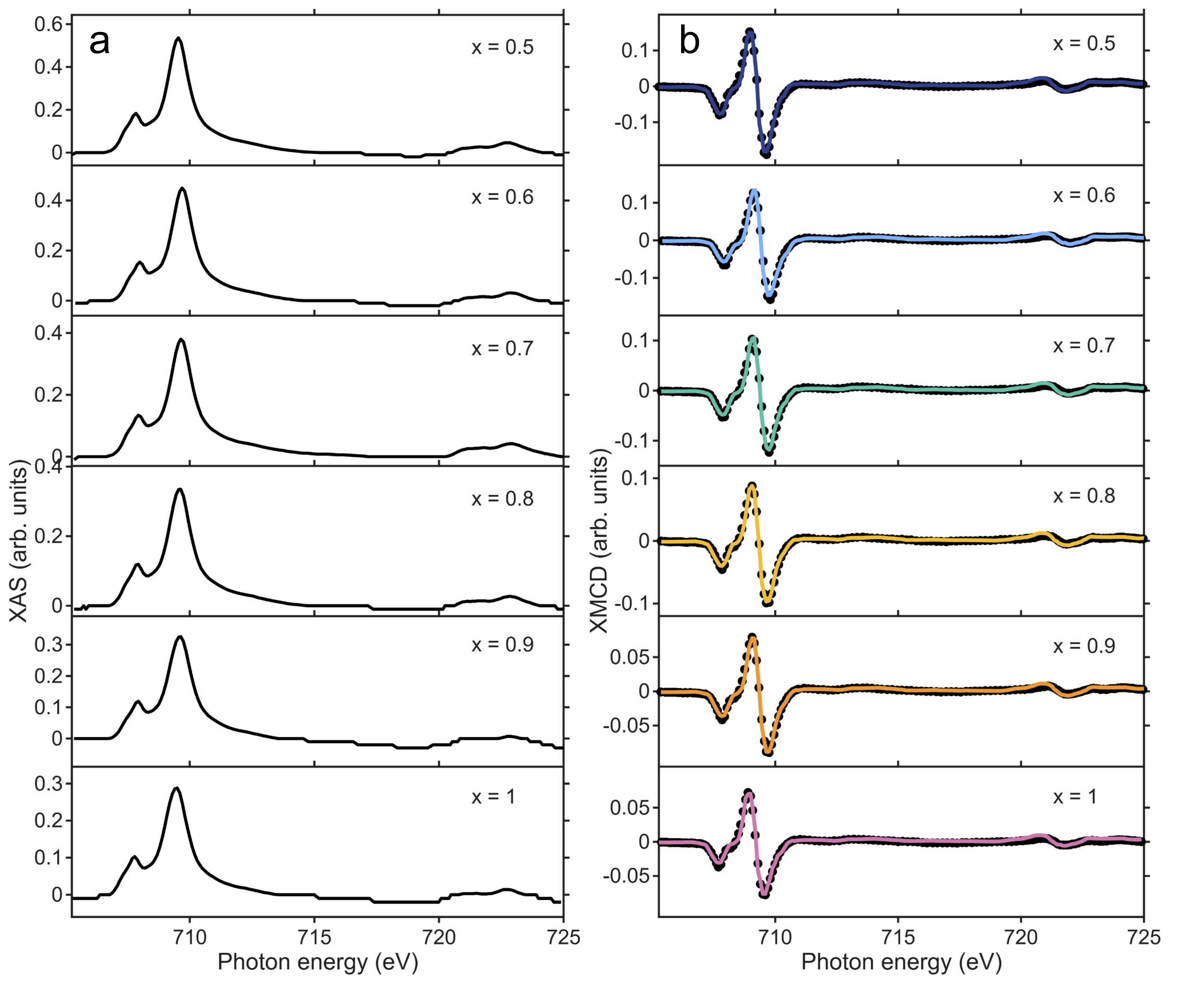}
    \caption{X-ray magnetic circular dichroism (XMCD) of Fe L$_{2,3}$ edges in LAFO x=0.5 to x=1.0 (black) fit to reference spectra (colors) of $\gamma$-Fe$_2$O$_3$ (62.5\% Fe$^{3+}_{Oh}$ 37.5\% Fe$^{3+}_{Td}$) \cite{brice-profeta_2005}, GaFeO$_3$  (100\% Fe$^{3+}_{Oh}$) \cite{Kim2006} as explained in the main paper }
    \label{fig:XMCD_sup}
\end{figure}

\section{X-Ray Absorption Spectroscopy and X-Ray Magnetic Circular Dichroism}
X-Ray Absorption Spectroscopy (XAS) and X-Ray Magnetic Circular Dicroism (XMCD) measurements were taken at the Fe L$_{2,3}$ edges to determine the Fe$^{3+}$ distribution among the octahedral and tetrahedral sites (Fig S1). The resulting XMCD scans are fit to experimental reference spectra with known iron cation distributions: $\gamma$-Fe$_2$O$_3$, GaFeO$_3$ \cite{brice-profeta_2005, Kim2006}. Cation distributions from these fits are listed in Table S2 to make Fig 3(c) in the main paper. 

\begin{table}[t]
\renewcommand{\thetable}{S\arabic{table}}
\caption{Octahedral and tetrahedral Fe$^{3+}$ cation distribution calculated from XMCD}
\label{tab:params}
\begin{ruledtabular}
\begin{tabular}{cccc}
Al Composition \ \ & Fe$^{3+}_{Oh}$ (\%) \ \  & \ \ Fe$^{3+}_{Td}$ (\%) \ \ & Error (\%) \\
\midrule
0.5 & 54.6 & 45.4 & 1.5  \\  
0.6 & 54.5  & 45.5 & 1.5 \\ 
0.7 & 55.1  & 44.9 & 1.6   \\
0.8 & 53.9  & 46.1 & 1.4  \\
0.9 & 54.3  & 45.7 & 1.1 \\
1.0 & 53.6  & 46.4 & 1.9   \\  
\end{tabular}
\end{ruledtabular}
\end{table}

\begin{figure}[t]
    \centering
    \renewcommand{\thefigure}{S\arabic{figure}}
    \includegraphics[width=\columnwidth]{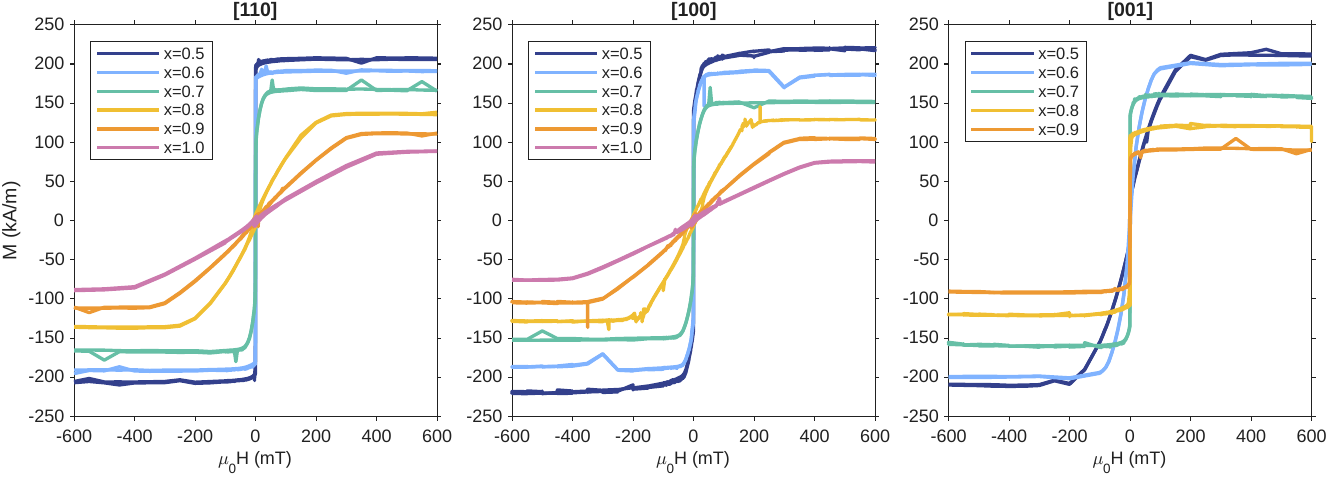}
    \caption{ Magnetization (M) dependence of applied field (H) for all LAFO compositions measured along (a) [110], (b) [100] and (c) [001] }
    \label{fig:squid_sup}
\end{figure}

\section{Static Magnetic Characterization}
Static magnetic characterization was performed using a Quantum Design Evercool SQUID magnetometer along two in-plane directions [110] and [100] and out-of-plane [001] for each composition of LAFO (Fig \ref{fig:squid_sup}). Each plot clearly shows a decrease in saturation magnetization with increasing Al composition. The in-plane [110] and [100] measurements show increasing saturation fields with increasing Al composition, while the out-of-plane [001] measurements show decreasing saturation fields with increasing Al. The values for average saturation magnetization and saturation fields along each crystallographic direction measured are listed in Table S3 which are used to create Fig 3 in the main paper. The transition from an in-plane easy direction to PMA can be clearly observed at LAFO x=0.7.  

\begin{table}[t]
\renewcommand{\thetable}{S\arabic{table}}
\caption{Saturation magnetization (M$_S$) and saturation field (H$_S$) values determined from SQUID data in Fig \ref{fig:squid_sup}. M$_S$ is averaged for each composition and H$_S$ is given for each chystalagraphic direction measured.}
\label{tab:params}
\begin{ruledtabular}
\begin{tabular}{ccccc}
Al Composition & M$_S$ (kA/m) & H$_S \parallel$  [110] (mT) & \ \ H$_S \parallel$ [100] (mT) \ \ & H$_S \parallel$ [001] (mT) \\
\midrule
0.5 & 217 $\pm$ 3 & 10 $\pm$ 6 & 30 $\pm$ 2 & 175 $\pm$ 6 \\  
0.6 & 195 $\pm$ 7 & 10 $\pm$ 6 & 29 $\pm$ 2 & 74 $\pm$ 9\\ 
0.7 & 164 $\pm$ 7 & 45 $\pm$ 6 & 67 $\pm$ 6 & 11 $\pm$ 2 \\
0.8 & 132 $\pm$ 7 & 192 $\pm$ 6 & 181 $\pm$ 1 & 15 $\pm$ 2\\
0.9 & 106 $\pm$ 5 & 281 $\pm$ 6 & 275 $\pm$ 2 & 15 $\pm$ 6\\
1.0 & 81  $\pm$ 10 & 349 $\pm$ 9 & 351 $\pm$ 6 & 13 $\pm$ 8 \\  
\end{tabular}
\end{ruledtabular}
\end{table}

\section{Broadband Ferromagnetic Resonance}
Broadband ferromagnetic resonance (FMR) measurements in a coplanar-waveguide-geometry were carried out for all Al doping compositions along the in-plane [100] and out-of-plane [001] directions. In Fig. S3, we plot $\Delta$H$_{hwhm}$ linewidth from FMR measurements along the in-plane [100] direction for LAFO x=0.5 and x=0.6 and along out-of-plane [001] direction for LAFO x=0.7 to x=1.0. We see low damping for each composition with a small increase in damping with increasing Al composition.

\begin{figure}[t]
    \centering
    \renewcommand{\thefigure}{S\arabic{figure}}
    \includegraphics[width=\columnwidth]{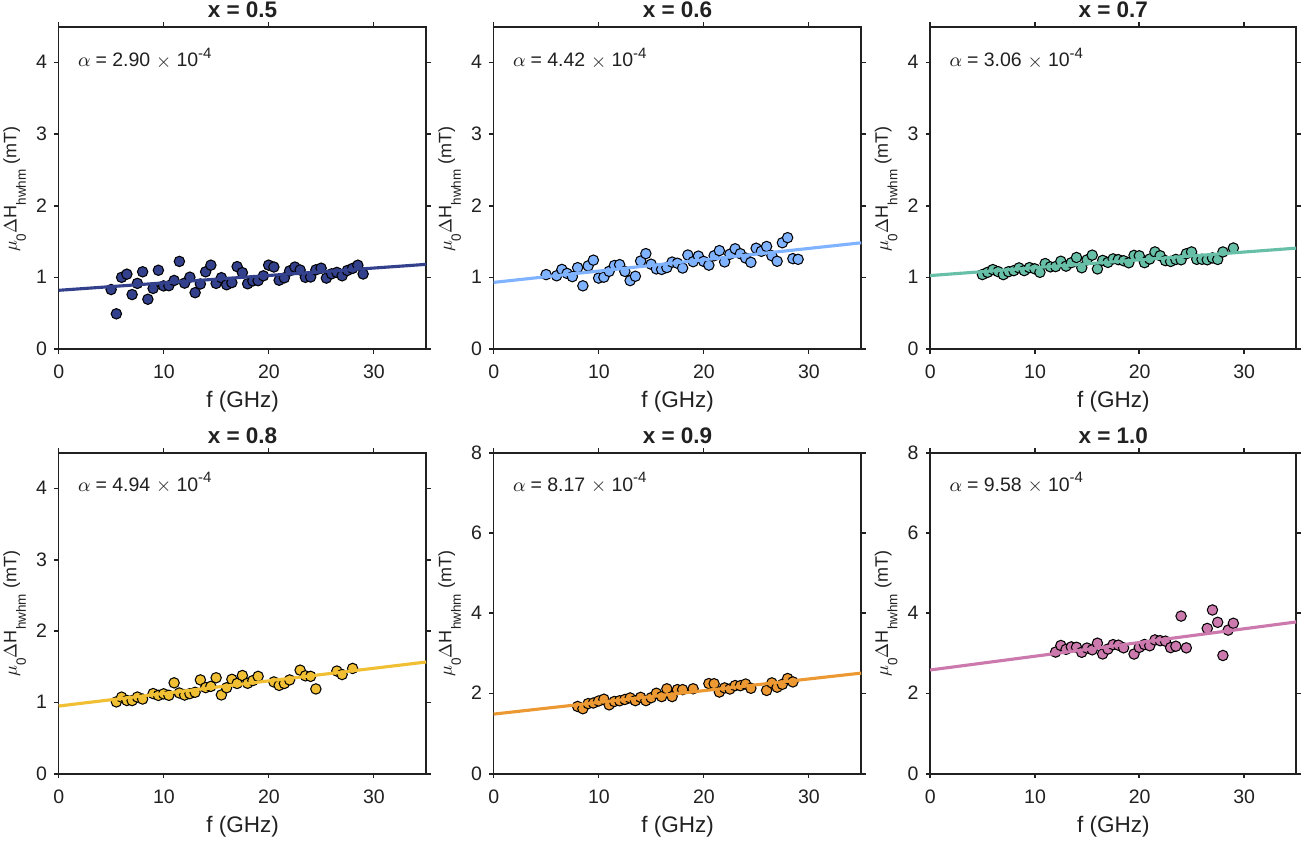}
    \caption{Frequency dependence of $\Delta$H$_{hwhm}$ linewidth for each LAFO composition measured along the easy plane}
    \label{fig:OOP_linewidth}
\end{figure}

\section{ST-FMR}
ST-FMR measurements in which a microwave charge current is accompanied by a dc current enable us to deduce a damping-like torque and field-like torque \cite{Karimeddiny2021-mo,Li2021-lf,Liu2011-vg,Nan2015-dq}. From the dc current dependence of the ST-FMR linewidth, we deduce the damping-like SOT from:

\begin{equation}
    |\theta_{DL}|=\frac{2|e|}{\hbar}\frac{(H_{res}+M_{eff}/2) \mu_0 M_S t_M}{|sin\phi|}\left|\frac{d\alpha_{eff}}{dJ_{dc}}\right|,
\end{equation}

where $\alpha_{eff}=|\gamma|\Delta H/ (2\pi f)$ and $|\gamma|$ is the gyromagnetic ratio of LAFO, $f$ is the microwave frequency (here we use 5 GHz for all measurements), t$_M$= 13 nm is the LAFO thickness and $\phi=45^\circ$, the in-plane angle between the magnetization and current, which is set to the value that returns maximum signal. Saturation magnetization M$_S$ is taken from SQUID measurements (Fig. 2 in main paper). M$_{eff}$ is deduced from FMR measurements and ranges from 200 mT in LAFO x=0.5 to -300 mT in LAFO x=1.0 (Fig. 4 in main paper). Data for the dc current dependence of linewidth with linear fits of the slope for each composition are shown in Fig. S4.   

Similarly, the field-like torque efficiency can in principle be determined from the dc current dependence of the FMR resonance field \cite{Liu2011-vg,Karimeddiny2021-mo}. However, the resonance field is significantly more sensitive to Joule heating than the linewidth. H$_{res}$ has been shown to follow close to the same T$^{3/2}$ dependence as $M_s$ does \cite{Kuanr2009-ra}. Therefore, heating induced by the applied dc current produces substantial nonlinearities in the resonance field shift. As a result, we are unable to reliably extract field-like torque efficiencies. %

\begin{figure}[t]
    \centering
    \renewcommand{\thefigure}{S\arabic{figure}}
    \includegraphics[width=\columnwidth]{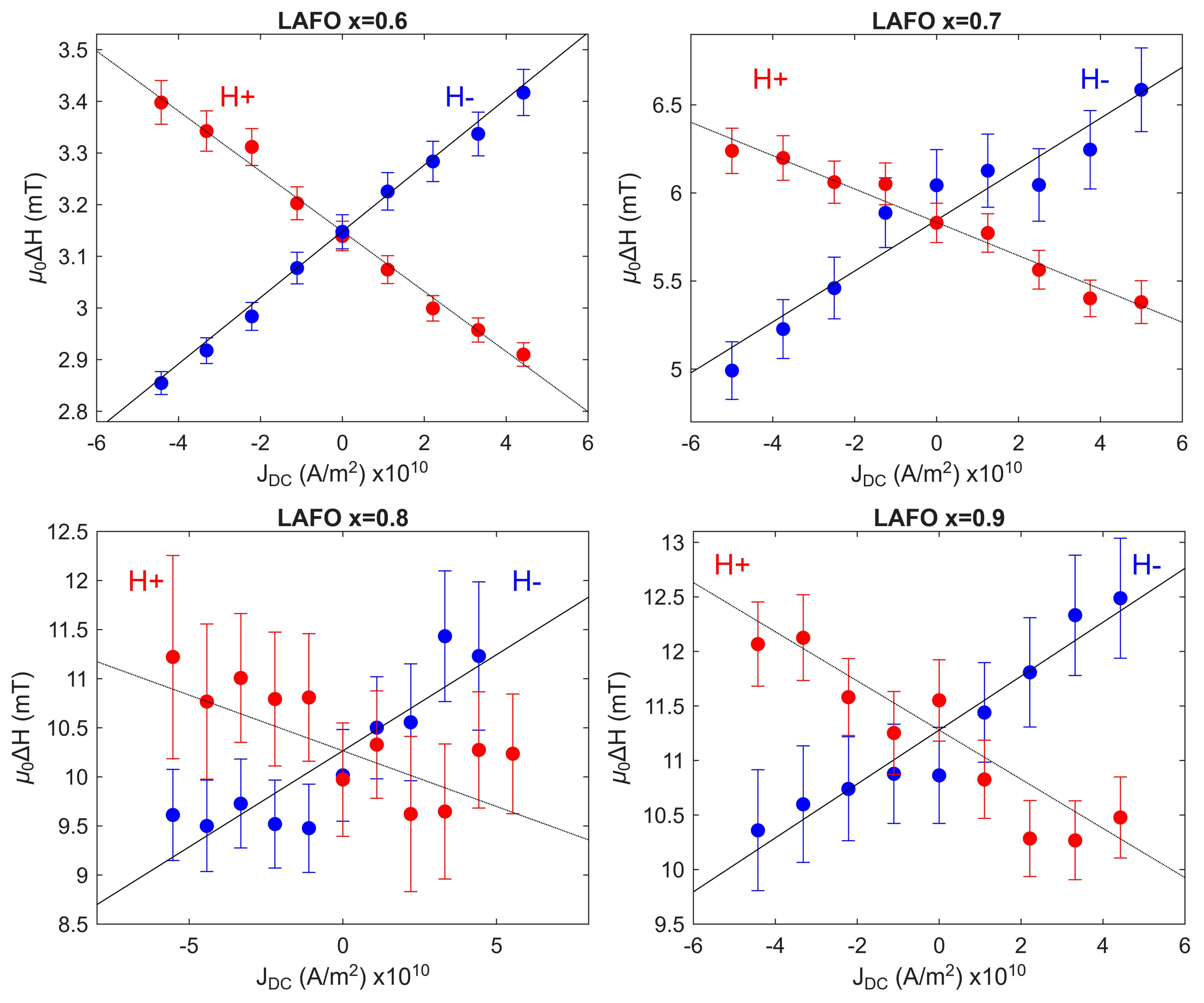}
    \caption{dc current density dependence of FMR linewidth for positive (red) and negative (blue) with applied fields for each composition with resolvable signal. Fits for each brach are shown in black.}
    \label{fig:OOP_linewidth}
\end{figure}


\section*{\label{sec:ref} References}
\bibliography{refs}